\documentclass[preprint,aps,12pt,preprintnumbers,eqsecnum,nofootinbib]{revtex4-2}
\usepackage{xcolor}
\usepackage{graphicx}
\usepackage{epstopdf} 
\usepackage{braket}
\usepackage{amssymb,amsmath,amsfonts,comment,latexsym,graphicx,epstopdf,float}
\usepackage{color}
\usepackage{fancyvrb}
\usepackage{chngcntr}
\usepackage{etoolbox}
\usepackage{ulem}
\usepackage{array}
\usepackage{slashed}
\usepackage{multirow}

\usepackage[colorlinks=true,citecolor=blue,linkcolor=red,filecolor=cyan,urlcolor=magenta]{hyperref}

\usepackage{float}

\newcommand{\be}{\begin{equation}}
\newcommand{\ee}{\end{equation}}
\newcommand{\ba}{\begin{eqnarray}}
\newcommand{\ea}{\end{eqnarray}}

\newcommand{\grts}{\raise.3ex\hbox{$>$\kern-.75em\lower1ex\hbox{$\sim$}}}
\newcommand{\lets}{\raise.3ex\hbox{$<$\kern-.75em\lower1ex\hbox{$\sim$}}}

{\catcode`\|=\active\gdef\Braket#1{\left<\mathcode`\|"8000\let|\bravert 
{#1}\right>}}

\def\bravert{\egroup\,\vrule\,\bgroup}

\usepackage{color}
\usepackage{amssymb,amsmath,amsfonts}
\usepackage{epstopdf} 
\usepackage{braket}

\usepackage{autobreak}

\begin{document}
%
%
\title{\vspace*{0.5in} 
Note on scattering in asymptotically nonlocal theories
\vskip 0.1in}
\author{Christopher D. Carone}\email[]{cdcaro@wm.edu}
\author{Mikkie R. Musser}\email[]{mrmusser@wm.edu}
\affiliation{High Energy Theory Group, Department of Physics,
William \& Mary, Williamsburg, VA 23187-8795, USA} 
%
%
\date{August 21, 2023}
\begin{abstract}
It is possible to formulate theories with many Lee-Wick particles such that a limit exists where the low-energy theory approaches the form of 
a ghost-free nonlocal theory.   Such asymptotically nonlocal quantum field theories have a derived regulator scale that is hierarchically smaller than
the lightest Lee-Wick resonance; this has been studied previously in the case of asymptotically nonlocal scalar theories, Abelian and non-Abelian 
gauge theories, and linearized gravity.   Here we consider the dependence on center-of-mass energy of scattering cross sections in these theories.  
While Lee-Wick resonances can be decoupled from the low-energy theory, scattering amplitudes nonetheless reflect the emergent nonlocality 
at the scale where the quadratic divergences are regulated.   This implies observable consequences in theories designed to address the hierarchy 
problem, even when the Lee-Wick resonances are not directly accessible.  
\end{abstract}

\maketitle
\newpage

\section{Introduction} \label{sec:intro}  
Quantum field theories with higher-derivative quadratic terms are of interest since these additional terms can lead to more convergent loop 
amplitudes.  This has an impact on the renormalizability of such theories and whether renormalization involves fine tuning.  If the highest power of  
derivatives $\partial_\mu\partial^\mu$ appearing in the quadratic terms is finite, then propagators will have additional poles.   Lee-Wick 
theories~\cite{LeeWick:1969,LeeWick:1972}, including the Lee-Wick Standard Model~\cite{Grinstein:2007mp} (see also \cite{Carone:2008iw}), have 
this feature and have been argued to be consistent with unitarity~\cite{Cutkosky:1969fq,Grinstein:2008bg,Anselmi:2017yux,Anselmi:2017lia,Anselmi:2018kgz} 
and macroscopic causality~\cite{Grinstein:2008bg}.  On the other hand, $\partial_\mu\partial^\mu$ may appear as the argument of an entire, 
transcendental function, so that the modified quadratic terms imply no additional propagator poles.   These are the ghost-free nonlocal theories that 
have met considerable attention in the recent literature~\cite{Tomboulis:1997gg,Modesto:2011kw,Biswas:2011ar,Ghoshal:2017egr,Buoninfante:2018mre,Ghoshal:2020lfd}.

It is possible to formulate another class of theories that interpolates between Lee-Wick theories and ghost-free nonlocal theories: these are the asymptotically nonlocal theories 
described in Refs.~\cite{Boos:2021chb,Boos:2021jih,Boos:2021lsj,Boos:2022biz}.   An asymptotically nonlocal theory is one of a sequence of finite-derivative theories that 
approaches a ghost-free nonlocal theory as a limit point.  We review the construction of asymptotically nonlocal theories in Sec.~\ref{sec:framework}.  In an ordinary Lee-Wick 
theory, the scale at which quadratic divergences are cancelled is set by the mass of the lightest Lee-Wick resonance. For example, if one were to decouple the Lee-Wick 
partners in the Lee-Wick Standard Model, fine-tuning in the Higgs boson squared mass would be reintroduced.   In asymptotically nonlocal theories, the Lee-Wick
partners may be heavy, while the light scalar mass is regulated by an emergent nonlocal scale $M_\text{nl}$, that is hierarchically smaller than the lightest Lee-Wick resonance 
mass, $m_1$,   
\begin{align}
M_\text{nl}^2 \sim {\cal O}\left(\frac{m_1^2}{N}\right) \, .
\end{align}
Here, $N$ is the number of propagator poles in a given theory, which provides a source of parametric suppression~\cite{Boos:2021chb,Boos:2021jih,Boos:2021lsj,Boos:2022biz}.   
Note that approaches to achieving a parametric suppression of the regulator scale have appeared in other contexts in the literature~\cite{Dvali:2007hz,Buoninfante:2018gce}. 

Asymptotically nonlocal theories have been explored previously in scalar theories~\cite{Boos:2021chb}, Abelian~\cite{Boos:2021jih} and non-Abelian gauge 
theories~\cite{Boos:2021lsj}, and in linearized gravity~\cite{Boos:2022biz}.   These papers discussed the higher-derivative and auxiliary field formulation of these 
theories (that is, equivalent theories in which higher-derivative terms are eliminated in favor of additional fields).  These papers also 
demonstrated the emergence of the nonlocal regulator scale in a variety of loop amplitudes, and in resolving gravitational singularities.    However, what was not 
considered was the implications for scattering cross sections.  For example, if asymptotically nonlocal theories interpolate between Lee-Wick and ghost-free
nonlocal theories, how is this transition reflected in the dependence of scattering cross sections on center-of-mass energy?   If propagators fall off exponentially with 
Euclidean momentum in asymptotically nonlocal theories, does one expect scattering amplitudes to grow without bound above the emergent nonlocal scale, given that 
the momentum transfers are not Euclidean?    (We will see later that the answer is no.)   In general, one 
expects that the physics associated with the emergent nonlocal regulator scale should also be apparent in scattering amplitudes as the Lee-Wick resonances 
are decoupled.  We explore this expectation in the present note by considering the momentum-dependence of an $s$-channel scattering cross section in an 
asymptotically nonlocal toy model that captures the qualitative features one expects to find in more realistic theories.  This fills a gap in the discussion that appeared in the previous literature~\cite{Boos:2021chb,Boos:2021jih,Boos:2021lsj,Boos:2022biz}. 

Our paper is organized as follows.  In Sec.~\ref{sec:framework}, we review the construction of asymptotically nonlocal theories, including our assumptions about 
how the limiting nonlocal theory is reached.  We define the model that we study later and discuss the form of the self-energy corrections to the propagator in the higher-derivative formulation of the theory.  Loop corrections encode the resonance widths, which truncate the growth in the scattering amplitudes that is associated with the emergent 
nonlocality.  In Sec.~\ref{sec:equiv}, we show in a simple example that the same results are obtained whether one works in the higher-derivative or (with more effort) in the 
auxiliary-field formulation of the theory.  In Sec.~\ref{sec:energy}, we describe how we implement mass and wave function renormalization in the higher-derivative theory and we 
present numerical results for the momentum dependence of the amplitudes that are of interest to us.  In Sec.~\ref{sec:conc}, we summarize our conclusions.   

\section{Framework and a toy model} \label{sec:framework} 
Our framework can be illustrated in a theory of a real scalar field $\phi$:  We seek a sequence of theories that approaches the nonlocal form
\begin{equation}\label{eq:lagexp}
\mathcal{L}_\infty = -\frac{1}{2} \phi \, (\Box + m_\phi^2)\,  e^{\ell^2 \Box} \, \phi -V(\phi) 
\end{equation}
as a limit point.  The exponential of the box operator shown in Eq.~(\ref{eq:lagexp}) is familiar from the literature on ghost-free nonlocal theories~\cite{Tomboulis:1997gg,Modesto:2011kw,Biswas:2011ar,Ghoshal:2017egr,Buoninfante:2018mre,Ghoshal:2020lfd}, and regulates loop integrals at the 
scale $1/\ell$.   A theory that approaches Eq.~(\ref{eq:lagexp}) when $N\rightarrow\infty$ is given by
\begin{equation}
\mathcal{L} = -\frac{1}{2}\phi \, (\Box+m_\phi^2) \left( 1+\frac{\ell^2\Box}{N-1} \right)^{N-1}\!\phi -V(\phi)~,
\end{equation}
However, the propagator in this theory contains an $(N-1)^\textrm{th}$ order pole, which has no immediate particle interpretation. We can remedy this by taking 
the $\ell_j$ to be nondegenerate,
\begin{equation}\label{eq:lagnondeg}
\mathcal{L}_N = -\frac{1}{2}\phi \, (\Box+m_\phi^2) \left[ \prod_{j=1}^{N-1}  \left( 1+\frac{\ell_j^2\Box}{N-1} \right)\right]\phi-V(\phi)~,
\end{equation}
which approaches the same limiting theory, Eq.~(\ref{eq:lagexp}), provided that $\ell_j$ approach a common value $\ell$ as $N$ becomes large.  For finite $N$, 
the propagator is given by
\begin{equation}
\label{eq:DF}
D_F(p^2) = \frac{i}{p^2-m_\phi^2} \prod_{j=1}^{N-1} \left( 1-\frac{\ell_j^2 p^2}{N-1} \right)^{-1}~.
\end{equation}
This has $N$ first-order poles; the massive states associated with the higher-derivative quadratic terms have masses $m^2_j \equiv (N-1)/\ell_j^2$.   The results of Refs.~\cite{Boos:2021chb,Boos:2021jih,Boos:2021lsj,Boos:2022biz} were not sensitive to how the $N \rightarrow \infty$,   $\ell_j \rightarrow \ell$ limit was reached.  
A convenient parameterization was given by
\begin{equation}\label{eq:masses}
m_j^2 = \frac{N}{\ell^2} \frac{1}{1-\frac{j}{2 N^P}} ~, \hspace{1cm} \textrm{for} ~j=1~.~.~.~N-1 \,\,\ , 
\end{equation}
for $P > 1$.   Away from the limit point, the propagator in Eq.~(\ref{eq:DF})  can be decomposed using partial fractions as a sum over simple poles with residues of 
alternating signs (a familiar outcome in higher-derivative theories~\cite{Pais:1950za}).  These correspond to an alternating tower of ordinary particles and ghosts.  We 
refer the reader to Ref.~\cite{Boos:2021chb} for the construction of an auxiliary field formulation that holds for arbitrary $N$.  We will discuss an auxiliary field 
formulation that is useful in the case where $N=2$ in Sec.~\ref{sec:equiv}.

\begin{figure}[t]
\includegraphics[width=.8\textwidth]{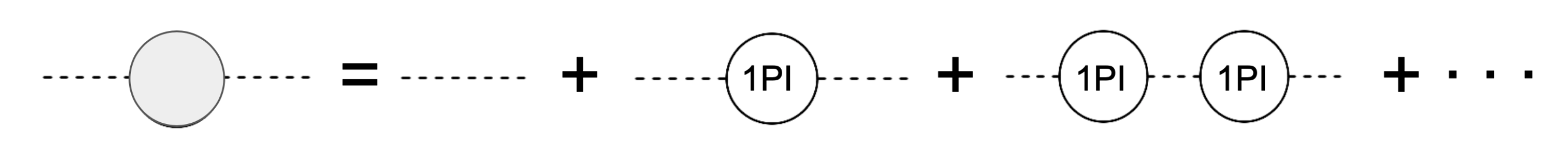}
\caption{Full $\phi$ propagator} \label{fig:prop}
\end{figure}

Writing the tree-level propagator in terms of the resonance masses $m_j$, one has
\begin{equation}
D_F(p^2)=\frac{i}{(p^2-m_\phi^2)\prod_{j=1}^{N-1}(1-p^2/m_j^2)}~,
\label{eq:propnice}
\end{equation}
The product in the denominator approaches a growing exponential for Euclidean momentum, which accounts for the better convergence properties of loop amplitudes discussed in  
Refs.~\cite{Boos:2021chb,Boos:2021jih,Boos:2021lsj,Boos:2022biz}.    To study the consequences of this form in scattering, and 
motivated by simplicity, we couple the $\phi$ field to complex scalar fields $\chi_a$, for $a=1\ldots 2$,
\begin{equation}
\label{eq:toy}
\mathcal{L}_{toy} = \mathcal{L}_N -\chi^{a*}\, (\Box+m_\chi^2) \, \chi^a -g\,\phi\, \chi^{a*}\chi^a~,
\end{equation}
where the summation on $a$ is implied, and we consider the $s$-channel scattering process $\chi_1 \chi_1 \rightarrow \chi_2 \chi_2$ in the center-of-mass frame.   We focus on $s$-channel processes as they are often associated with large momentum transfers in 
realistic theories at colliders, and they provide a relatively direct way to study the energy-dependence implied by the distinctive form of the propagators found in asymptotically nonlocal theories.   We expect that the example we study will provide a qualitative understanding of $s$-channel processes in these theories, independent of the precise choice of fields appearing on the external lines or the spin of the exchanged particle.  The product in the 
denominator of Eq.~(\ref{eq:propnice}) approaches an exponential that rapidly decreases as a function of the squared center-of-mass energy $s$, above the 
emergent nonlocal scale $M_\text{nl}\equiv 1/\ell$.
What prevents arbitrary growth of the propagator is the widths of the resonances (just as 
it would be had we chosen, for example, $s=m_\phi^2$).   To capture that physics, we define the one-particle irreducible self-energy function
$-i M^2( p^2)$ and compute the full propagator shown in Fig.~\ref{fig:prop}.
The diagrammatic resumation gives
\begin{equation}
\label{eq:resummed}
D_F^\text{full} =\frac{i}{(p^2-m_\phi^2)\prod_{j=1}^{N-1}(1-p^2/m_j^2)-M^2(p^2)}~,
\end{equation}
which reduces to the familiar expression~\cite{Peskin:1995ev} when $N=1$, where the product is replaced by the identity.  We will see in Sec.~\ref{sec:energy} that the imaginary part of $M^2(p^2)$ limits the maximum value of the scattering amplitude.

\section{Equivalent approaches} \label{sec:equiv} 
Before considering the implications of the momentum dependence of Eq.~(\ref{eq:resummed}), we briefly digress to consider the general form of this expression.  In an auxiliary field formulation
of the higher-derivative theory, the higher-derivative terms are eliminated in favor of additional fields (each corresponding to a propagator pole).  In that theory, there are a number of possible one-particle irreducible self energy diagrams, depending on the choice of external lines.   Here we look at the scattering process $\chi_1 \chi_1 \rightarrow \chi_2 \chi_2$ in the auxiliary theory in the simplest case of $N=2$ and show how the loop corrections conspire to exactly reproduce the corrected form of the higher-derivative propagator in Eq.~(\ref{eq:resummed}).   We expect this  to hold for arbitrary $N$ on general grounds; however, this example illustrates how computations that are easy in the higher-derivative form of the theory can become prohibitively complicated in its auxiliary form.  Hence, in Sec.~\ref{sec:energy}, we return to working with the higher-derivative theory.

In the case where $N=2$, the Lagrangian is given by
\begin{equation}
{\cal L} = -\frac{1}{2} \hat{\phi} \, (\Box+m_\phi^2) (1 + \ell^2 \Box) \, \hat{\phi} + {\cal L}_\text{int} \,\,\, ,
\label{eq:n2lag}
\end{equation}
where the Lee-Wick partner to the particle with mass $m_\phi$ has mass $M \equiv 1/\ell$, and ${\cal L}_\text{int}$ contains the coupling to the $\chi$ fields. We assume $M>m_\phi$. We place a hat on the field that
appears in the higher-derivative form of the theory for later notational convenience.  An equivalent theory
can be identified using an auxiliary field $\widetilde{\phi}$:
\begin{equation}
{\cal L}=-\frac{1}{2} \left(1+\frac{m_\phi^2}{M^2}\right) \hat{\phi} \, \Box \, \hat{\phi} - \widetilde{\phi} \, \Box \, \hat{\phi} +\frac{1}{2} M^2 \widetilde{\phi}^2 -\frac{1}{2} m_\phi^2 \, \hat{\phi}^2
+ {\cal L}_\text{int} \,\,\, .
\label{eq:auxform}
\end{equation}
The $\widetilde{\phi}$ is non-dynamical and can be eliminated from the generating functional for the theory by performing the corresponding Gaussian functional integral.   Operationally, 
the resulting Lagrangian is what one obtains from Eq.~(\ref{eq:auxform}) by replacing $\widetilde{\phi}$ using its equation of motion,
\begin{equation}
\widetilde{\phi} = \frac{1}{M^2} \Box \hat{\phi} \,\,\, .
\end{equation}
With this substitution, one recovers Eq.~(\ref{eq:n2lag}).  It is convenient to rescale $\hat{\phi} = \xi^{-1} \hat{\phi}_1$ and $\widetilde{\phi} = \xi \widetilde{\phi}_1$, with
\begin{equation}
\xi \equiv \left(1+\frac{m_\phi^2}{M^2}\right)^{1/2} \,\,\, ,
\end{equation}
so that
\begin{equation}
{\cal L}=-\frac{1}{2} \hat{\phi}_1 \, \Box \, \hat{\phi}_1 - \widetilde{\phi}_1 \, \Box \, \hat{\phi}_1 +\frac{1}{2} M^2 \xi^2 \widetilde{\phi}_1^2 -\frac{1}{2} \,  \xi^{-2} m_\phi^2 \, \hat{\phi}_1^2
+ {\cal L}_\text{int} \,\,\, .
\end{equation}
Shifting $\hat{\phi}_1 = \phi_1 - \widetilde{\phi}_1$ leads to the following form:
\begin{equation}
{\cal L} = - \frac{1}{2} \Phi^T \Box \, {\cal K} \Phi -\frac{1}{2} \Phi^T {\cal M} \, \Phi + {\cal L}_\text{int} \, \,
\end{equation}
where 
\begin{equation}
\Phi \equiv \left(\begin{array}{c} \phi_1 \\  \widetilde{\phi}_1 \end{array}\right) \,\, , \,\,\,\,\,
{\cal K} = \left(\begin{array}{cc} 1 & 0 \\ 0 & -1 \end{array}\right) \,\,\,\, \mbox{ and } \,\,\,\,\,
{\cal M} = \xi^{-2} m_\phi^2 \, \left(\begin{array}{ccc} 1 & \hspace{1em}& -1 \\ -1 && 1 - \xi^4 \frac{M^2}{m_\phi^2} \end{array}\right) \,\,\, .
\end{equation}
The kinetic matrix ${\cal K}$ has the form one expects in a Lee-Wick theory, with one field having a canonically normalized, but wrong-sign kinetic term.  The mass squared 
matrix ${\cal M}$ is off-diagonal.  A transformation of the form $\Phi = R \, \Phi_0$ with
\begin{equation}
R = \left(\begin{array}{cc} \cosh\theta & \sinh\theta \\ \sinh\theta & \cosh\theta \end{array}\right)
\end{equation}
will leave ${\cal K}$ unchanged but can be used to diagonalize ${\cal M}$.  We find that this is the case for
\begin{equation}
\theta = \frac{1}{2} \ln \! \left(\frac{M^2-m_\phi^2}{M^2+m_\phi^2}\right) \,\,\, ,
\end{equation}
which leads to the simple form
\begin{equation}
R = \frac{1}{\sqrt{M^4 - m_\phi^4}} \left(\begin{array}{ccc} M^2 &\hspace{1em}& -m_\phi^2 \\ -m_\phi^2 && M^2 \end{array} \right) \,\,\, .
\end{equation}.  
In terms of the mass eigenstate fields $\Phi_0$, the Lagrangian becomes
\begin{equation}
{\cal L} = -\frac{1}{2} \Phi_0^T \left( \Box \, {\cal K}_0 + {\cal M}_0\right) \Phi_0  + {\cal L}_\text{int} \,\,\, ,
\end{equation}
where 
\begin{equation}
\Phi_0 \equiv \left(\begin{array}{c} \phi_0 \\  \widetilde{\phi}_0 \end{array}\right) \,\, , \,\,\,\,\,
{\cal K}_0 = \left(\begin{array}{cc} 1 & 0 \\ 0 & -1 \end{array}\right) \,\,\,\, \mbox{ and } \,\,\,\,\,
{\cal M}_0 = \left(\begin{array}{ccc} m_\phi^2 & \hspace{1em}& 0 \\ 0 && -M^2 \end{array}\right) \,\,\, .
\label{eq:alldiag}
\end{equation}
This result reproduces the same propagator poles expected in the higher-derivative theory, Eq.~(\ref{eq:n2lag}).

The field redefinitions that led to Eq.~(\ref{eq:alldiag}) allow us to rewrite $\hat{\phi}$ in terms of the mass eigenstate fields
\begin{equation}
\hat{\phi} = \frac{M}{\sqrt{M^2-m_\phi^2}} \,  [ \phi_0 - \widetilde{\phi}_0 ] \,\,\, .
\end{equation}
The interaction assumed in our toy model, shown in Eq.~(\ref{eq:toy}), then becomes
\begin{equation}
{\cal L}_\text{int} = -g  \frac{M}{\sqrt{M^2-m_\phi^2}} v_0^T \Phi_0 \, \chi^* \chi \,\,\, ,
\label{eq:newvertex}
\end{equation}
where we define $v_0^T \equiv (1,-1)$.   In the $\Phi_0$ field basis, the self-energy function can be written as a two-by-two matrix, 
$-i M^2(p^2)_{\alpha\beta}$, where the indices represent either $ \phi_0$ or  $\widetilde{\phi}_0$.   The full matrix propagator takes the form
\begin{equation}
D_F^\text{full}(p^2) = i \left[ p^2 {\cal K}_0 -{\cal M}_0 - M^2(p^2) \right]^{-1} \,\,\, .
\label{eq:matprop}
\end{equation}
However, the self-energy matrix in Eq.~(\ref{eq:matprop})  can be expressed in terms of the self-energy function that appears in the higher-derivative theory:
\begin{equation}
M^2(p^2)_{\alpha\beta} = [v_0 v_0^T]_{\alpha\beta} \frac{M^2}{M^2-m_\phi^2} \, \hat{M}^2(p^2) \,\,\,.
\label{eq:mmhat}
\end{equation}
One can understand Eq.~(\ref{eq:mmhat}) as follows:  the one-$\chi$-loop amplitude following from Eq.~(\ref{eq:newvertex}) is the same as in the higher-derivative 
theory, up to the prefactors appearing in Eq.~(\ref{eq:mmhat}).   Higher-loop contributions may involve additional internal $\chi$ loops, as well as $\phi_0$ and $\widetilde{\phi}_0$ internal
lines, where the latter will always appear together and re-sum to give the higher-derivative propagator for $\hat{\phi}$.  Hence, the function $\hat{M}^2(p^2)$ is diagrammatically the 
same as the one appearing in the higher-derivative theory.   Using the vertex in Eq.~(\ref{eq:newvertex}) the Feynman amplitude for the $s$-channel process
$\chi_1 \chi_1 \rightarrow \chi_2 \chi_2$ is given by
\begin{equation}
i {\cal A}(\chi_1 \chi_1 \rightarrow \chi_2 \chi_2) = \frac{- i \, g^2 M^2}{M^2-m_\phi^2} \, v_0^T \left[p^2 {\cal K}_0 -{\cal M}_0 - [v_0 v_0^T] \frac{M^2}{M^2-m_\phi^2} \, \hat{M}^2(p^2) \right]^{-1} v_0 \,\, .        
\label{eq:matrix2into2}
\end{equation}
With the matrix structure of Eq.~(\ref{eq:matrix2into2}) completely specified, one may evaluate the inverse and simplify.  One finds
\begin{equation} 
i {\cal A}(\chi_1 \chi_1 \rightarrow \chi_2 \chi_2) = -g^2 \frac{ i }{(p^2 - m_\phi^2)(1-p^2/M^2) - \hat{M}^2(p^2)} \,\,\, ,      
\end{equation}
which precisely reproduces the form expected in the higher-derivative formulation, following from Eq.~(\ref{eq:resummed}), when $N=2$.  For larger $N$, it is clearly preferable to work directly with the
higher-derivative form of the loop-corrected propagator, avoiding the field redefinitions and other avoidable algebra that was illustrated by this example.  We use 
the higher-derivative approach in the section that follows.

\section{Energy dependence of amplitudes} \label{sec:energy} 

To evaluate a scattering amplitude that contains the full propagator in Eq.~(\ref{eq:resummed}), we must adopt 
an explicit form for the self-energy function $M^2(p^2)$.  In the theory presented in Eq.~(\ref{eq:toy}), one finds at one-loop using dimensional regularization\footnote{Alternatively, one could use a cut off regulator with the on-shell renormalization scheme discussed later, with no effect on the results.} that 
\begin{equation}
\begin{split} \label{eq:explicitM}
M^2(p^2)&=-\frac{n_\chi g^2}{16\pi^2} \int_0^1 dx \left[ \frac{2}{\epsilon} -\gamma +\textrm{ln}4\pi -\textrm{ln}\frac{\Delta}{\mu^2} \right] \\
&= -\frac{n_\chi g^2}{16\pi^2}\left[ \int_0^1 dx \left[ \frac{2}{\epsilon} -\gamma +\textrm{ln}4\pi -\textrm{ln}\frac{|\Delta|}{\mu^2} \right] +i\pi \sqrt{1-\frac{4m_\chi^2}{p^2}} \, \Theta(p^2-4\, m_\chi^2)\right]~,
\end{split}
\end{equation}
where $n_\chi=2$ is the number of $\chi$ fields, $\Delta \equiv m_\chi^2-x(1-x) \, p^2$ and $\Theta$ is the Heaviside step function.  For 
$p^2>4 \, m_\chi^2$, the self-energy has an imaginary part, which approaches a constant value when $p^2 \gg m_\chi^2$.   The logarithmic divergence in Eq.~(\ref{eq:explicitM}) is absent in physical quantities after mass and wave function renormalization.\footnote{One could alternatively consider the possibility that the $\chi$-sector is asymptotically nonlocal, which would lead to a much more cumbersome, but finite, one-loop self-energy function.  Such a complication is unnecessary for the present study.}
\begin{figure}[b]\label{fig:selfenergy}
\includegraphics[width=0.5\textwidth]{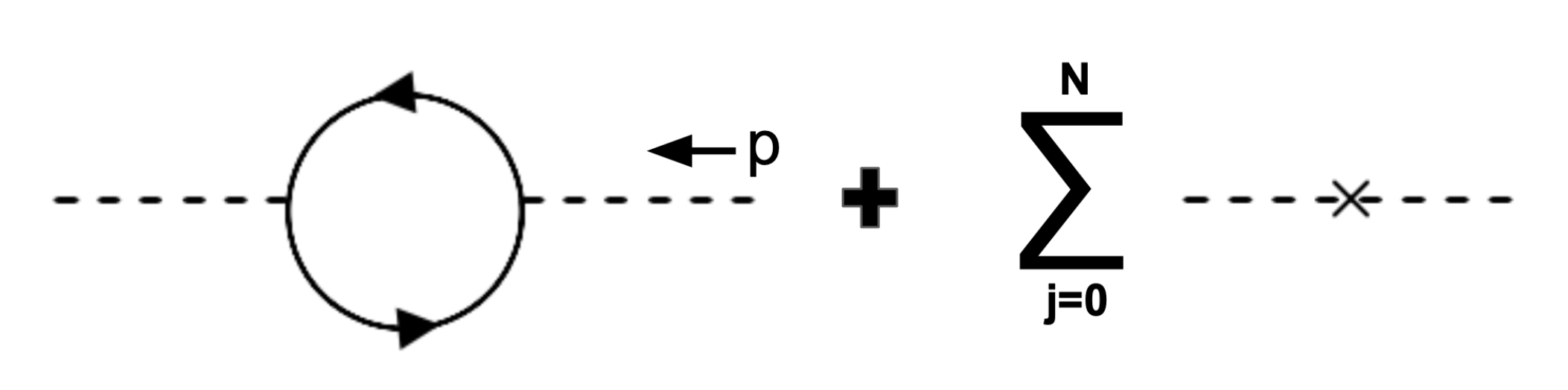}
\caption{One-loop diagram corresponding to the $\phi$ self-energy with counterterms proportional to $a_k \, p^{2k}$.}
\end{figure}
Since the quadratic operator in our theory takes the form of a polynomial in $\Box$, as can be seen in Eq.~(\ref{eq:lagnondeg}), we can define our renormalized theory as
\begin{equation}\label{eq:withct}
\mathcal{L}_N = -\frac{1}{2} \phi \,(\Box +m^2_\phi)  \left[ \prod_{j=1}^{N-1}\left( 1+\frac{\Box}{m_j^2} \right) - \sum_{k=0}^{N} \delta_k \, \Box^k \right]\phi  -V(\phi)~,
\end{equation}
where $m_\phi$ and the $N-1$ masses $m_j$ are physical masses and the $\delta_k$ correspond to counterterms that will be 
determined by renormalization conditions.  It follows from Eq.~(\ref{eq:withct}) that the renormalized propagator is
\begin{equation}\label{eq:rendfull}
D^\text{full}_F=\frac{i}{(p^2-m_\phi^2)\prod_{j=1}^{N-1}(1-p^2/m_j^2)-M_r^2(p^2)}~,
\end{equation}
where 
\begin{equation}
M_r^2(p^2) = M^2(p^2) - \sum_{k=0}^N a_k \, p^{2k}~,
\end{equation}
with $a_k\equiv -\delta_k (-1)^k$.  The coefficients $a_k$ need to be fixed by $N+1$ renormalization conditions.    Taking into account that Lee-Wick particles are unstable, 
we require that the location of the propagator poles on the real axis correspond to the physical masses $m_j$, giving us $N$ conditions
\begin{equation}\label{eq:poleloc}
\textrm{Re}\,M_r^2(m_j^2)=0~, \hspace{1cm} j=0\ldots N-1,
\end{equation}
where $m_0 \equiv m_\phi$. Fixing the wave function renormalization of the $\phi$ field in the higher-derivative theory gives us the remaining condition.  As $\phi$ is only relevant for internal lines in the diagrams of interest to us, there are no problems introduced by leaving the wave
function renormalization at any pole non-canonical, as no compensating factors need to be introduced in the scattering amplitudes of interest.  Hence, we make a convenient choice for the remaining condition
\begin{equation}
\textrm{Re}\, M_r^2(0)=0~.
\label{eq:zcond}
\end{equation}
Note that this is equivalent to identifying $g$ as the physical coupling defined  at the reference point $p^2=0$.
Eq.~(\ref{eq:zcond}) determines the coefficient $a_0$ which absorbs the divergent part of Eq.~(\ref{eq:explicitM}):
\begin{equation}
a_0 = -\frac{n_\chi g^2}{16\pi^2} \left[ \frac{2}{\epsilon} -\gamma +\textrm{ln}4\pi -\textrm{ln}\frac{m_\chi^2}{\mu^2} \right]~.
\end{equation}
With this choice, the remaining conditions, Eq.~(\ref{eq:poleloc}), may be written
\begin{equation} \label{eq:polecon}
\textrm{Re}\, M_r^2(m_j^2) = \frac{n_\chi g^2}{16\pi^2} \int_0^1 dx~ \textrm{ln}\left(\frac{| m_\chi^2-x(1-x)m_j^2 |}{m_\chi^2}\right) - \sum_{k=1}^N a_k \, m_j^{2k}=0~,
\end{equation}
for $j=0\ldots N-1$.   These $N$ equations allow one to solve for the remaining coefficients and therefore the full loop-corrected 
propagator.   Defining $\textrm{Re}\, M_r^2(m_j^2) = \tilde{M}^2(m_j^2)- \sum_{k=1}^N a_k \, m_j^{2k}=0$, we may write
Eq.~(\ref{eq:polecon}) in matrix form
\begin{equation}
\left(\begin{array}{c} \tilde{M}^2(m_0^2) \\ \tilde{M}^2(m_1^2) \\ \vdots \\ \tilde{M}^2(m_{N-1}^2) \end{array}\right) = 
\left(\begin{array}{cccc} m_0^2 & m_0^4 & \ldots & m_0^{2N} \\
m_1^2 & m_1^4 & \ldots & m_1^{2N}  \\
\multicolumn{4}{c}{\vdots} \\
m_{N-1}^2 & m_{N-1}^4 & \ldots & m_{N-1}^{2N} \end{array}\right) \,
\left(\begin{array}{c} a_1 \\ a_2 \\ \vdots \\ a_N \end{array}\right)  \,\,\, ,
\end{equation}
or more compactly, $\tilde{M}_i = m_{ij} \, a_j$.  Hence, the desired coefficients may be computed numerically by evaluating
\begin{equation}
a_k = [m^{-1}]_{kj} \, \tilde{M}_j \,\,\, .
\end{equation}
Note that the $a_k$ for $k=1 \ldots N$ are independent of $1/\epsilon$ and represent finite radiative corrections that vanish when $g \rightarrow 0$.

As discussed earlier, we focus on the $s$-channel scattering process $\chi_1\chi_1\rightarrow\chi_2\chi_2$.  The choice of different $\chi$ 
fields in the initial and final state eliminates $t$- and $u$-channel diagrams, which do not affect our qualitative
conclusions but would complicate the discussion.  We plot the $s$-dependence of the scattering amplitudes for both 
$P=1$ and $P=1.5$, where $P$ is the parameter appearing in Eq.~(\ref{eq:masses}).   Note that for $P>1$, Eq.~(\ref{eq:masses}) implies
that all the $m_j$ approach a common value as $N \rightarrow \infty$.  This is not the case for $P=1$ when $j$ is of order $N$.  
However, it was found in Ref.~\cite{Boos:2021chb} that even in this case loop amplitudes approach the asymptotically 
nonlocal form, with Euclidean loop momenta exponentially suppressed above an emergent nonlocal scale.  Hence, we 
present this case here as well.   

\begin{figure}[t]
\includegraphics[width=\textwidth]{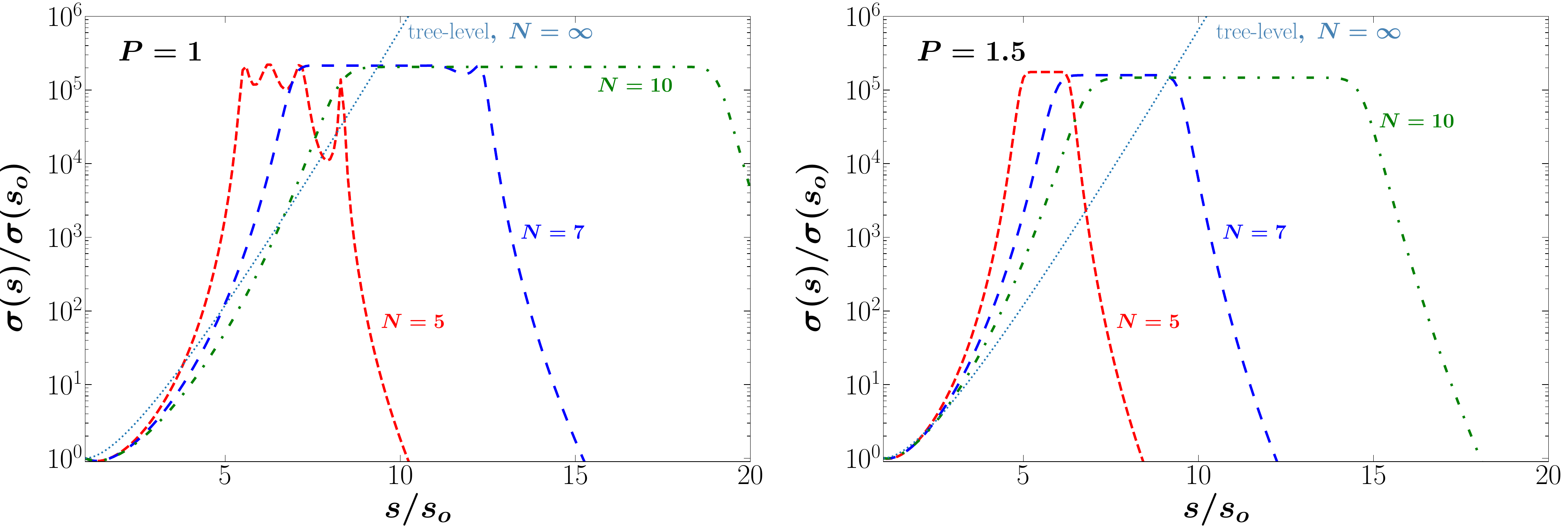}
\caption{Dependence of the scattering cross section for $\chi_1\chi_1\rightarrow\chi_2\chi_2$ with the squared center of mass energy $s$, normalized
to the cross section at $s_0=M_\text{nl}^2$ for the values of $N$ and $P$ shown.  In units where $M_\text{nl}=1$, this example corresponds
to the choices $g=1$, $m_\phi=0.01$ and $m_\chi = m_\phi/4$.
\label{fig:plots1}}
\end{figure}
Results for the scattering cross section, for a number of choices for $N$ (the total number of poles), and for $P=1$ and $P=1.5$ are shown in Fig.~\ref{fig:plots1}.   The cross section results are normalized to their values when $\sqrt{s}$ is set equal to the nonlocal scale, {\it i.e.}, $s = M_\text{nl}^2$.     We see that the results for $P=1$ and $P=1.5$ are qualitatively similar.  The cross section plots have a region in $s$ immediately above the nonlocal scale where the cross section grows, with the growth gradually approaching the exponential form expected in the nonlocal limiting theory as $N$ becomes large.  The cross section levels off in the resonance region above the mass of the first Lee-Wick particle, with hints of resonants peaks visible at the smaller values of $N$ and $P$, due to the smaller overlap between adjacent resonances.   We do not expect the product in 
Eq.~(\ref{eq:rendfull}) to approximate an exponential as the resonance region is approached for two reasons: (1) mathematically, the product 
deviates from its exponential limiting form as $s$ increases at finite $N$, and (2) this rapidly decreasing term is eventually surpassed by the contribution from the self-energy term as $s$ increases.   Above the resonance region, the result falls off as the square of the highest power of  momentum in the polynomial that appears in the propagator denominator.  Our numerical results in Fig.~\ref{fig:plots1} are consistent with these expectations.  We also
note that the normalization factor $\sigma(s_0)$ asymptotes to a constant as $N$ becomes large, so the results shown do not hide any uncontrolled growth or suppression.\footnote{For example, in units where $M_\text{nl}=1$ and $s_0 =M_\text{nl}^2$, we find numerically that $\sigma(s_0)=a_0+b_0/N +c_0/N^2 + {\cal O}(1/N^3)$, with $a_0 = 17.187$, $b_0=7.826$ and $c_0=5.526$, in the case where $P=1$, assuming the other parameter choices given in the caption of Fig.~\ref{fig:plots1}.}

\begin{figure}[t]
\includegraphics[width=\textwidth]{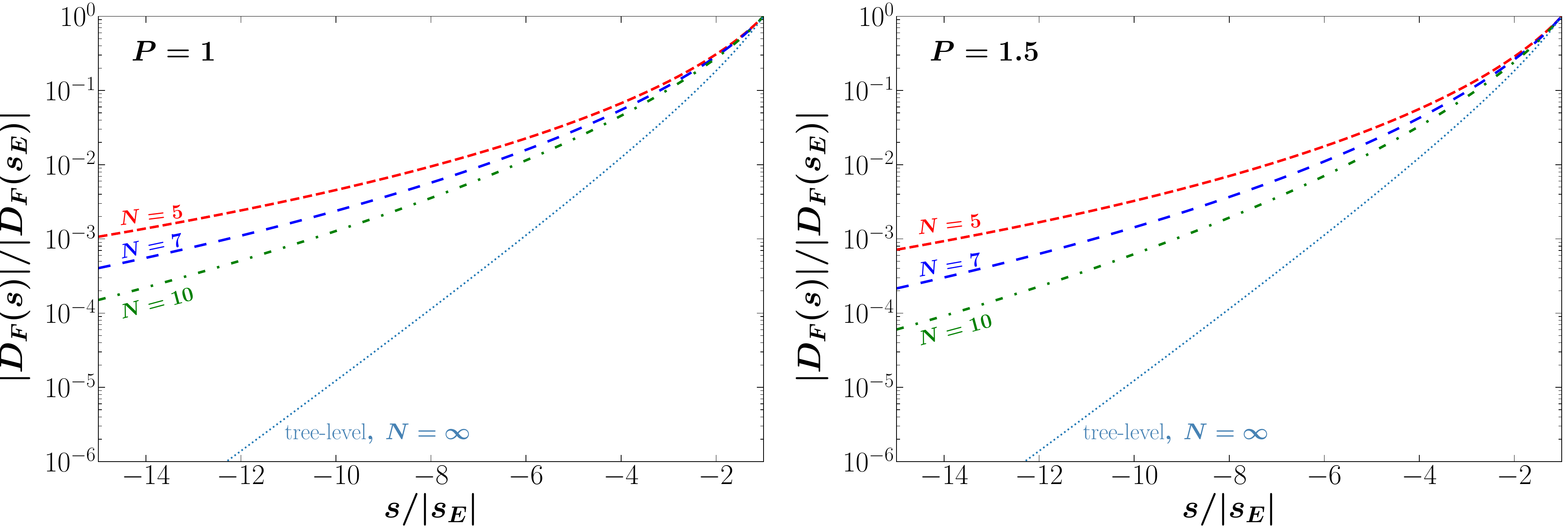}
\caption{Dependence of the magnitude of the full propagator, at one loop, for Euclidean $s$, normalized to the same quantity evaluated at $s_E=-M_\text{nl}^2$ for the values of $N$ and $P$ shown.  In units where $M_\text{nl}=1$, this example corresponds
to the choices $g=1$, $m_\phi=0.01$ and $m_\chi = m_\phi/4$.
\label{fig:plots2}}
\end{figure}

Previous work on asymptotic nonlocality focused on loop amplitudes where momentum is Euclidean after Wick rotation.  At higher-loop order, the full propagator may appear within other loops, 
which motivates us to check the behavior of Eq.~(\ref{eq:rendfull}) for Euclidean momentum.  In Fig.~\ref{fig:plots2}, we plot the magnitude of the propagator for Euclidean values of the $s$-channel 
momentum (normalized to the same quantity evaluated at $s = - M_\text{nl}^2$), as a point of comparison.  We see that the results monotonically decrease with increasing $|s|$ and
approach the exponential form of the limiting theory with 
increasing $N$.  This is qualitatively consistent with the behavior encountered in the study of loop amplitudes in Refs.~\cite{Boos:2021chb,Boos:2021jih,Boos:2021lsj}.

Finally, it is interesting to note that in the cross section examples we present in Fig.~\ref{fig:plots1}, the range in $\sqrt{s}$ that we consider is relatively small, a factor of at most $\sim 4.5$ between the smallest and 
largest values.  Yet  within this range, one can see an energy dependence for the scattering cross section  that differs from what one might expect to find in either a simple Lee-Wick theory or a 
ghost-free nonlocal theory.   This may make these class of theories phenomenologically distinguishable from the other two in realistic theories, 
at least in the case where it is possible experimentally to probe the relevant range of center-of-mass energies.

\section{Conclusions} \label{sec:conc} 
Asymptotically nonlocal theories are a sequence of Lee-Wick theories that approach a ghost-free nonlocal theory in their low-energy limit~\cite{Boos:2021chb,Boos:2021jih,Boos:2021lsj,Boos:2022biz}.  
The nonlocal modification of the quadratic terms that is obtained in the limiting theory suggests that a derived nonlocal regulator scale will emerge in theories with a finite number 
of Lee-Wick particles, as the appropriate limit is approached.   This regulator scale is hierarchically smaller than the mass of the lightest Lee-Wick resonance, and its emergence has been 
explored in past work on scalar field theories~\cite{Boos:2021chb}, gauge theories~\cite{Boos:2021jih,Boos:2021lsj} and in linearized gravity~\cite{Boos:2022biz}.    The regulator appears 
because the nonlocal form factor in the limiting theory provides a suppression factor for Euclidean momentum, and hence a faster fall-off in the Wick-rotated propagators that appear in 
loop diagrams.  For simple scattering processes, where momentum transfers are not Euclidean, one may worry that the effect of the form factor is to cause all scattering amplitudes to 
diverge.  This is not the case, for the same reason that propagators are not infinite when the center-of-mass energy sits exactly at a resonance value: the growth is limited by the resonance width.   
In the present case, we take the resonance widths into account by including the self-energy in the propagator, working in the higher-derivative form of the theory for arbitrary $N$.   We showed in the simple case where $N=2$ that the same results are obtained whether one formulates the problem in the higher-derivative or Lee-Wick forms of the theory, where the latter exchanges 
higher-derivative terms for additional fields; however, the higher-derivative form is easier to work with as the number of propagator poles $N$ becomes large.

With the self-energy included in the propagator in an $s$-channel scattering process in a simple toy model, we identified mass 
and wave function renormalization 
conditions and explored how the propagator behaves as moved towards the asymptotically nonlocal limit; we considered the case where the squared 
momentum transfer $s$  flowing through the propagator is positive (relevant for scattering) or negative (relevant for loop amplitudes due to Wick rotation).   For $s>0$ we found that cross sections 
will grow above the nonlocal scale, will plateau in the region of Lee-Wick resonances, and then fall off at $s$ larger than the heaviest resonance.   The region of growth gradually approaches an exponential form as $N$ increases and 
the maximum is determined by the imaginary part of the self-energy in the higher-derivative theory.  On the other hand, for $s<0$, one finds monotonic suppression 
as $|s|$ becomes large, with the magnitude of the propagator approaching a dying exponential in the same way.\footnote{In fact, one can show that the deviation of the finite-$N$ result from the exponential limiting form is what one would expect for an exponential that is approximated by a product, as in
Eq.~(\ref{eq:lagnondeg}).} This is consistent with the behavior that leads to an 
emergent regulator scale in loop amplitudes discussed in our earlier work~\cite{Boos:2021chb,Boos:2021jih,Boos:2021lsj,Boos:2022biz}.  

The growth of cross sections with center-of-mass energy followed by a broad resonant plateau and then subsequent fall off is neither the qualitative behavior of a simple Lee-Wick theory nor a 
ghost-free nonlocal theory; this is not surprising since the model we study interpolates between the two.   Qualitatively, the first signs of growth in the cross section due to emergent 
nonlocality might not look very different  at a collider experiment (assuming a realistic theory) from what one might expect from the tail of a heavy resonance whose mass is just outside 
an experiment's kinematic reach.   Since such heavy resonances are not observed, the bounds on the emergent nonlocality scale are likely in the multi-TeV range.  An exact bound would require a 
dedicated collider analysis in a realistic theory, which may be of interest for future work.
\vspace{-1em}
\begin{acknowledgments} 
We thank the NSF for support under Grant PHY-2112460.  We thank Jens Boos for his comments on the manuscript.
\end{acknowledgments}


\end{document}